\documentstyle[aps,prb,epsf,epsfig]{revtex}
\begin{document}
\pagestyle{plain}
\draft
\title{
The electronic structure of CuSiO$_3$ -- a possible candidate for
a new inorganic spin-Peierls compound ?}
\author{H.~Rosner, S.-L.~Drechsler, K.~Koepernik, R.~Hayn, and H.~Eschrig}
\address{Institut f\"ur Festk\"orper- und Werkstofforschung Dresden,
P.O.~Box 270016, D-01171 Dresden, Germany}
\date{\today}
\twocolumn[\hsize\textwidth\columnwidth\hsize\csname@twocolumnfalse\endcsname
\maketitle

\begin{abstract}
Electronic structure calculations are presented for the well-known
CuGeO$_3$ and the recently discovered isostructural CuSiO$_3$
compounds.  The magnitude of the dispersion in chain direction is
considerably smaller for CuSiO$_3$, whereas the main interchain
couplings are rather similar in both compounds.  Starting from 
extended one-band tight-binding models fitted to the bandstructures, the
exchange integrals were estimated for both compounds in terms of a
spatially anisotropic Heisenberg model.  Remarkable frustrating second
neighbor couplings  are
found both for intra- and inter-chain interactions.  A magnetic moment
of about 0.35 $\mu_B$ is predicted for CuSiO$_3$ in the N\'eel state.

\begin{pacs}
\end{pacs}
\end{abstract}] 

Low-dimensional spin systems such as chains or ladders are of
fundamental interest for contemporary solid state physics due to their
peculiar electronic and magnetic properties. During the last years,
many related materials have been found within the cuprate
family, famous for the high temperature superconductivity.  All
cuprates contain CuO$_4$ plaquettes.  In most cases it is
energetically favorable to connect these plaquettes by the formation
of chains or planes. According to the number ($n=1,2$) of oxygen atoms
shared by adjacent plaquettes, these compounds can be classified as
so-called edge-shared ($n=2$) or corner-shared ($n=1$) compounds.

Obviously, the type of sharing affects strongly the physical
properties of the compounds under consideration.  For example, corner
sharing leads to strong antiferromagnetic coupling between neighboring
plaquettes compared with the weak inter-chain
interactions.\cite{rosner97} As a result, the straight CuO$_3$ chain
in Sr$_2$CuO$_3$ is the best known realization of the one-dimensional
spin-1/2 Heisenberg model,\cite{tsvelik95} with an in-chain exchange
coupling of about 2200~K, but with a N\'eel temperature of only 5 K
and with an extremely small ordered magnetic moment of about 0.06~$\mu
_{\text{B}} $,\cite{kojima97} both due to a small residual interchain
exchange coupling. Spin-charge separation in the
excitation spectra could be observed for Sr$_2$CuO$_3$ and for the
double chain compound SrCuO$_2$.\cite{kim96}

Somewhat surprisingly, in contrast to the similarity between different
corner-shared chain compounds, the magnetic properties in the
edge-shared chain family exhibit a remarkable variance. Thus, the
edge-shared CuO$_2$ plaquettes in Li$_2$CuO$_2$ order
antiferromagnetically with a ferromagnetic arrangement along those
chains and with a large ordered moment of 0.9~$\mu
_{\text{B}}$,\cite{sapina90} whereas the same chain in CuGeO$_3$ shows
a spin-Peierls transition at low temperatures.\cite{hase93}
Antiferromagnetically ordered chains were observed in
Cu$_{1-x}$Zn$_x$GeO$_3$ for small concentrations of Zn
impurities.\cite{hase93a} It is noteworthy that, even for the
intensively studied CuGeO$_3$, a consensus with respect to the
quantitative description of competing or complementary interactions
such as the inter-chain coupling, frustration and spin-phonon coupling
has not been reached so
\vbox{
\begin{figure}[ht]
\epsfysize=9cm
\centerline{\epsfbox{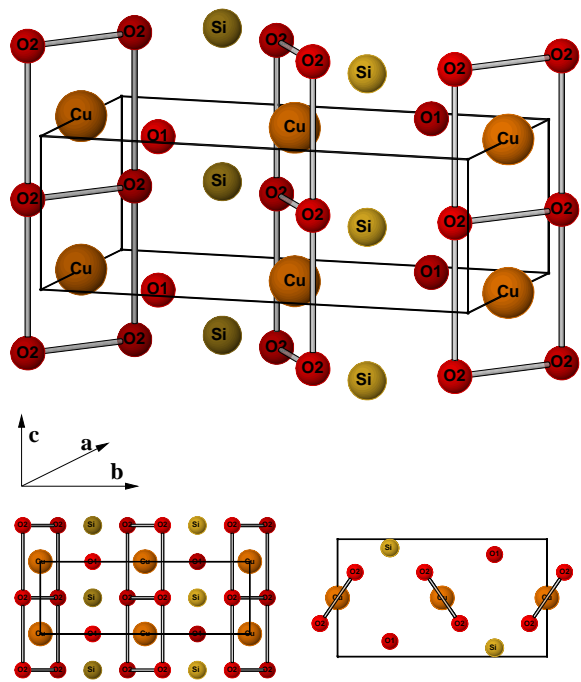}}\vspace{.3cm}
\caption{The orthorhombic unit cell of the CuSiO$_3$-crystal, perspective view
(top), front view (down left) and top view (down right).  The
edge-shared cuprate-chains run along the {\bf c} direction and are
canted against each other.}
\label{struct}
\end{figure}
}
\vskip-.75\baselineskip

\noindent
far,\cite{uhrig97,bouzeras99} despite the
achieved qualitative understanding of their influence on different
magnetically ordered states.

Naturally, the magnetic properties depend very sensitively on the
electronic interactions in these systems. Therefore, a comparative
study of the electronic properties of closely related systems can shed
light on the interactions responsible for the magnetically ordered
states mentioned above.  In this context, the recent discovery and
first investigations of the long searched for compound
CuSiO$_3$,\cite{otto99} which is isostructural to the prototypical
inorganic spin-Peierls system CuGeO$_3$ is of great scientific
interest.  The crystal structure of CuSiO$_3$ is shown in
Fig.~\ref{struct}. The most important feature for the magnetic
properties are the planar edge-shared CuO$_2$ chains running along
{\bf c}-direction. These chains are very similar to those of
CuGeO$_3$. The Cu-O(2) bond length in CuSiO$_3$ (CuGeO$_3$) is
1.941~\AA\ (1.942~\AA), the Cu-O(2)-Cu bonding angle is 94$^\circ$
(99$^\circ$).

Thus, the question arises, whether the very recently observed phase
transition \cite{baenitz00} near 8 K does point to a new inorganic
spin-Peierls system or to another ordered state realized at low
temperature.  To get theoretical insight into possible scenarios, we
present here comparative band-structure calculations and tight-binding
examinations for CuSiO$_3$ and CuGeO$_3$. In this context we note that
for the latter compound several (non full-potential) bandstructure
calculation have been reported (e.g. in
Ref.~\onlinecite{mattheiss94}), but to our knowledge the inter-chain
interaction has not been analyzed in detail.

The relevant electronic structure of these materials is very sensitive
to details of hybridization and charge balance. In order to obtain a
realistic and reliable hopping part of a tight binding Hamiltonian,
band-structure calculations were performed using the full-potential
nonorthogonal local-orbital minimum-basis scheme
\cite{koepernik99} within the local density approximation
(LDA). In the scalar relativistic calculations we used the exchange
and correlation potential of Perdew and Zunger.\cite{perdew81}
Cu($4s$, $4p$, 3$d$), O(2$s$, 2$p$, 3$d$), Ge(3$d$, 4$s$, 4$p$, 4$d$)
and Si(2$p$, 3$s$, 3$p$, 3$d$) states, respectively, were chosen as
minimum basis set. All lower lying states were treated as core
states. The inclusion of Ge 3$d$ and Si 2$p$ states in the valence
states was necessary to account for non-negligible core-core
overlaps. The O and Si 3$d$ as well as the Ge 4$d$ states were taken
into account to increase the completeness of the basis set. The
spatial extension of the basis orbitals, controlled by a confining
potential \cite{eschrig89} $(r/r_0)^4$, was optimized to minimize the
total energy.

The results of the paramagnetic calculation\cite{remark3} for
CuSiO$_3$ (see Fig.~\ref{band} (a)) and CuGeO$_3$ (see Fig.~\ref{band}
(b); we find similar results as the non full-potential calculation of
Ref.~\onlinecite{mattheiss94}) show a valence band complex of about 10
eV width with two bands crossing the Fermi level in both cases. These
two bands are well separated from the rest of the valence band complex
and show mainly Cu 3$d$ and O(2) 2$p$ character in the analysis of the
corresponding partial densities of states (not shown). We note that
the occupancy of the two O(2) 2$p$ orbitals along and perpendicular to
the chain (lying in the plaquette-planes) is rather different, but it
is almost identical for the corresponding orbitals in both compounds.
Therein, we found only a small admixture of O(1) 2$p$ and Ge 4$s$ and
4$p$ states, respectively, with a total amount of few percent. The
examination of the eigenstates of the latter bands at high symmetry
points yields an antibonding character typical for cuprates.  Here
these relatively narrow antibonding bands are half-filled. Therefore,
strong correlation effects can be expected which explain the
experimentally observed insulating groundstate.
Despite almost perfect qualitative one to one correspondence of
all valence bands and main peak structures in the densities of states
(DOS) (compare right panels in Fig.~\ref{band}), the most important
differences between both compounds 
\vbox{
\begin{figure}[ht]
\epsfysize=14.4cm
\centerline{\epsfbox{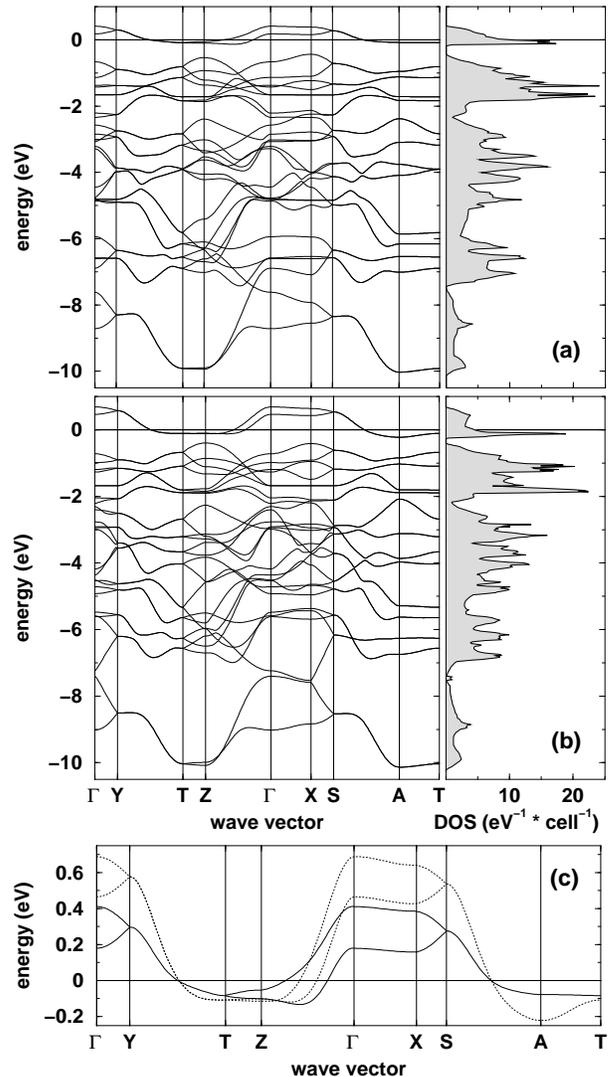}}\vspace{.3cm}
\caption{\label{band}
Band structure and total density of states for CuSiO$_3$ (a),
CuGeO$_3$ (b), and the zoomed antibonding bands (c) (CuSiO$_3$ full
lines, CuGeO$_3$ dashed lines). The Fermi level is at zero energy. The
notation of the symmetry points is as follows: Y = (010), T = (011), Z
= (001), X = (100), S = (110), A = (111). The chain direction
corresponds to Y--T, Z--$\Gamma$ and S--A.}
\end{figure}
}
\vskip-.25\baselineskip

\noindent
occur for the antibonding bands (shown in detail in
Fig.~\ref{band}(c)).  Therefore, we restrict ourselves to the extended
tight-binding analysis and the discussion of these antibonding bands.

The dispersion of these bands has been analyzed in terms of nearest
neighbor transfer (NN), next nearest neighbor transfer (NNN) and
higher neighbor terms in chain direction, but only NN hopping and a
diagonal transition term between the CuO$_2$-chains have been
considered (see Fig.~\ref{skizze}).  Then, the corresponding
dispersion relation takes the form
\vbox{
\begin{figure}
\begin{center}
\begin{minipage}{5.9cm}
\epsfxsize=\hsize\epsfbox{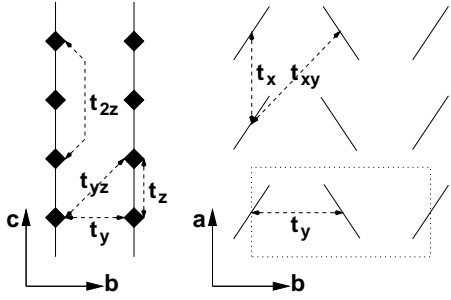}\vspace{.5cm}
\end{minipage}
\end{center}
\vspace{-2mm}
\caption{Schematical chain and stack arrangement of CuO$_2$-plaquettes,
respectively, and considered transfer processes within the bc-plane
(left panel) and in the ab-plane (right panel).}
\label{skizze}
\vspace{-3mm}
\end{figure}
} 
\begin{eqnarray}
E(\vec{k})&=&-2\big(\sum_{m=1,4}t_{mz}\cos (mz)
+\cos(x)\left[t_{x} +2t_{xz}\cos(z)\right]\nonumber \\
&&+\cos(y/2)\left[t_y+
2t_{yz}\cos(z)+2t_{xy}\cos(x)\right]\big),
\end{eqnarray}
where $x=k_za$, $y=k_yb$, $z=k_zc$.  Notice that in our effective
one-band description the upper band (see Fig.~\ref{band}(c)) e.g.\
along $\Gamma$--X corresponds to $k_y$ = 0, whereas the lower one
corresponds to $k_y$ = $2\pi /b$.  The assignment of the parameters
has been achieved by two numerically independent procedures: By
straightforward least square fitting of the whole antibonding band in
all directions and by using the bandwidths, the slopes and the
curvatures at special selected high symmetry points. The latter
procedure has the advantage to be less affected by hybridization
effects from lower lying bands near the bottom of the antibonding band
(being of some relevance near the Z-point in Fig.~\ref{band}).

\vbox{
\begin{table}
{\begin{tabular}{|c|c|c|c|c|c|c|}
&
t$ _{z} $&
t$ _{2z} $&
t$ _{3z} $&
t$ _{x} $&
t$ _{y} $&
t$ _{yz}$\\
\hline 
CuGeO$_3$&
-175 &
-51&
-5.5&
-20&
-34.1&
-20.6\\
CuSiO$_3$&
-88&
-31&
-4.5 &
-2.4&
-36 &
-21.2\\
\end{tabular}\par}\vspace{.5cm}
\caption{ \label{tabel1}
Transfer integrals $t_i$ (in meV) of the extended one-band
tight-binding model for CuGeO$_3$ and CuSiO$_3$. The remaining omitted
terms from Eq.~(1) were found to be irrelevant.}
\end{table}
}

\noindent 
The results are shown in Tab.~\ref{tabel1}. The errors can be
estimated between 1\% for the large and 10\% for the small parameters
from the difference of both mentioned above fitting procedures.  The
analyzed antibonding bands of both compounds exhibit a rather similar
shape except near the Z-points, where the hybridization with lower
lying bands produces an additional band-crossing for CuGeO$_3$ (see
Fig.~\ref{band}(c)).  Recall that the main difference to the
corner-shared chains as e.g.\ in Sr$_2$CuO$_3$ is a much smaller
in-chain NN transfer due to the different geometry.

In spite of the qualitative similarity, the calculated values for the
transfer integrals are quite different.  The in-chain dispersion is
nearly twice as large for CuGeO$_3$ in comparison to CuSiO$_3$. This
can be attributed mainly to the larger Cu-O-Cu bond angle in CuGeO$_3$
(99$^\circ$ and 94$^\circ$, respectively). However, this geometrical
effect is somewhat reduced by the different on-site energies of the
oxygen orbitals along and perpendicular to the chain (lying in the
plaquettes planes). The latter difference is reflected by the larger
separation of the corresponding bands at the Z-point in CuSiO$_3$ (see
Fig.~\ref{band}).

The inter-chain dispersions in $b$ direction are comparable. For both
compounds, we find also rather significant diagonal hopping terms
$t_{yz}$ which are reflected by different dispersions along the X--S
and the T--Z directions.  Somewhat surprisingly, we found a sizeable
dispersion in $x$-direction for CuGeO$_3$ but only a very weak one for
the CuSiO$_3$ counterpart.

From the transfer integrals discussed above, we conclude that both
compounds are not so well-defined quasi one-dimensional systems
as compared to the corner-shared CuO$_3$ chain compounds
\cite{rosner97,rosner99}.  The inter-chain coupling is rather
significant for CuGeO$_3$, and CuSiO$_3$ can even be regarded as an
anisotropic two-dimensional system.  Since
increasing inter-chain coupling tends
to destabilize the spin-Peierls
state\cite{inagaki83b}, a N\'eel ordered antiferromagnetic ground
state might be expected for CuSiO$_3$ in contrast to the spin-Peierls
state realized in CuGeO$_3$.

The obtained transfer integrals enables us to estimate the relevant
exchange integrals $J$. This knowledge is crucial for the derivation
and examination of magnetic model Hamiltonians of the spin-1/2
Heisenberg type
frequently used in the literature: 
\begin{equation}
H_{spin}={\sum_{ij}}^{\prime}J_{ij}\vec{S_i}\vec{S_j} \, .
\end{equation}
In general, the total exchange $J$ can be divided into an
antiferromagnetic and a ferromagnetic contribution $J$ = $J^{AFM} +
J^{FM}$.  In the strongly correlated limit, valid for typical
cuprates, the former can be calculated in terms of the one-band
extended Hubbard model $J^{AFM}_{ij}$ = $4t^2_{ij}/(U-V_{ij})$. The
indices $i$ and $j$ correspond to nearest and next nearest neighbors,
$U$ is the on-site Coulomb repulsion and $V_{ij}$ is the inter-site
Coulomb interaction. From experimental data \cite{parmigiani96} mapped
from the standard $pd$-model onto the one-band description, one
estimates $U - V$ $\sim$ 4.2 eV. For the sake of simplicity, we
neglect the difference in the quantity  $U - V$ in the compounds. The
calculated values for the exchange integrals are given in
Tab.~\ref{tabel2}.

\vbox{
\begin{table}
{\begin{tabular}{|c|c|c|c|c|c|c||c|c|}
&
J$ _{1}^{AFM} $&
J$ _{1} $&
J$ _{2} $&
J$ _{x} $&
J$ _{y} $&$J_{yz}$&$\mu_{th}$&$\mu_{exp}$\\
\hline 
CuGeO$_3$&
29&
15 &
2.5&
0.4 &
1.11 &0.4&0.17&0.21\\
CuSiO$_3$&
7.4&
3.8&
0.9&
0.006 &
1.25 &0.43&0.35&unknown\\
\end{tabular}\par}\vspace{.3cm}
\caption{\label{tabel2}
Exchange parameters $J_i$ (in meV) for CuGeO$_3$ and CuSiO$_3$, and
local magnetic moments (in $\mu_{B}$) in the N\'eel state derived
from them (see text). The experimental value $\mu_{ex}$ is an average
over various studies mentioned in the text.}
\vspace{-2mm}
\end{table}}

\noindent
The value of the NN exchange integral $J^{AFM}_{1}$
$\sim$ 30 meV in CuGeO$_3$ exceeds the experimental values of about 11
meV from inelastic neutron scattering data \cite{regnault96}, about 14
meV from magnetic susceptibility\cite{fabricius98} and about 22 meV
from Raman scattering \cite{kuroe97}.  This points to a significant
ferromagnetic contribution due to the
Goodenough-Kanamori-Anderson-type interaction\cite{anderson59}.
In the following, 
we shall adopt 15 meV for the resulting total
exchange coupling $J_1$ as a representative value, suggested by the
average of the above mentioned experimental data.  Owing to the lack
of experimental data we assume
the same ratio $J_1/J_1^{AFM}$ in CuSiO$_3$ as in CuGeO$_3$, suggested
by the quite similar O(2) 2$p$ orbital occupancies mentioned above.
For the latter compound, we note the reasonable agreement with the
available experimental data and most of our calculated
antiferromagnetic values for the remaining exchange parameters.
Hence, further possible ferromagnetic contributions seem to be less
relevant and are neglected in the following considerations.

Further simplification can be obtained mapping J$_1$
and the frustrated NN term J$_2$ onto an effective intra-chain coupling
$J_{\parallel}=J_1 -  1.12J_2$.\cite{fledderjohann97}  The
calculated values for $J_{\parallel}$ are 12.2 meV for CuGeO$_3$
and 2.8 meV for CuSiO$_3$, respectively.  The latter value is close to
the value of 2~meV reported by Baenitz {\it et al.} from a
one-dimensional fit of
magnetic susceptibility data.\cite{baenitz00}
We find also a considerable inter-chain frustration $J_{yz}=\beta J_y$
with $\beta$=0.36 (0.34) for the Ge- (Si-) compound.  This is in good
agreement with the suggestions of Uhrig \cite{uhrig97} $\beta \approx
0.5$ for CuGeO$_3$. 

Transfering the above mentioned idea to map frustrating terms onto one
effective coupling,\cite{fledderjohann97} we adopt
$J_{\perp}=J_y-2J_{yz}$ for the effective inter-chain exchange
parameters in $b$-direction.  The factor of two is introduced to
account approximately the twice as large number of second neighbors. The
effective anisotropy ratio $R = J_{\perp}/J_{\parallel}$ 
measures approximately the magnitude of quantum fluctuations.  In the
crossover region between one and two dimensions, quantum fluctuations
do strongly affect the magnitude of the staggered magnetization $m$
and the local Cu moment $\mu=g_{L}n_dm$ at $T=0$ for a N\'eel ground
state, where $g_L$=2.06 to 2.26 \cite{honda96} denotes the
(anisotropic) Land\'e-factor (tensor) for Cu$^{2+}$ in CuGeO$_3$ and
$n_d\approx $ 0.8 is the hole occupation number of the related Cu 3$d$
plaquette orbital.  Using the expression
\begin{equation}
m=0.39\sqrt{R}(1+0.095R)\ln^{1/3}(1.3/R) ,
\end{equation}
taken from Ref.~\onlinecite{sandvik99}, we arrive at 0.17$\mu_B$ in
reasonable agreement \cite{remark2} with the neutron data 0.22$\pm 0.02$
\cite{hase96} and 0.2 \cite{sasago96} for the disorder induced N\'eel
state achieved below 4.5K in Zn-doped CuGeO$_3$. The same approach
predicts a significantly larger value of about 0.35$\mu_B$ for
CuSiO$_3$ realized in a possible N\'eel state.

To summarize, our LDA-FPLO calculation reveals valuable insight into
the relevant couplings of CuGeO$_3$ and CuSiO$_3$.  We can classify
CuGeO$_3$ as a quasi one-dimensional compound with significant
inter-chain interaction, whereas CuSiO$_3$ is closer to an anisotropic
two-dimensional compound.  The significantly reduced energy scale of
the in-chain exchange interactions and the large inter-chain
interaction in CuSiO$_3$ are less favorable for a spin-Peierls state
than for a N\'eel order. However, due to the large frustrations other
states such as a spin-Peierls state cannot be excluded. Further
investigations are required to elucidate the unknown ground state.

We acknowledge fruitful discussions with M.~Baenitz, C.~Geibel,
W. Pickett and G.~Uhrig. This work was supported by individual grants
of the DAAD (H.R.) and the DFG (S.D.).\\[-.7cm]


\begin{thebibliography}{10}
\bibitem{remark3} For the tight-binding analysis, the paramagnetic solutions
are sufficient. Magnetic LSDA solutions were also found for both
compounds showing insulating behavior, but due to the insufficient
treatment of correlation in the LSDA the calculated gap is 
much too small.
\bibitem{remark2} The estimated moment should be regarded as a lower
bound due to the neglected weak interaction in $x$-direction which
reduces further the quantum fluctuations. Due to the
Mermin-Wagner-theorem, this interaction is crucial for the finite
N\'eel temperature not considered here.
\end{thebibliography}



\end{document}